\documentclass[letterpaper,twocolumn,10pt]{article}
\usepackage{usenix-2020-09}
\usepackage{amsmath,amsfonts}
\usepackage{algorithmic}
\usepackage{algorithm}

\usepackage{array}
\usepackage{textcomp}
\usepackage{stfloats}
\usepackage{url}
\usepackage{verbatim}
\usepackage{graphicx}
\usepackage{cite}
\usepackage{pbox}

\usepackage{xcolor}
\usepackage{hyperref}
\hypersetup{
    colorlinks=true,
    linkcolor=blue,
    filecolor=blue,      
    urlcolor=blue,
    pdftitle={ISF 671/22},
    pdfpagemode=FullScreen,
    citecolor=blue
    }
\usepackage{listings}
\usepackage{caption}
\usepackage{subcaption}
\usepackage{comment}

\usepackage{lscape}
\usepackage[nohyperlinks,printonlyused,withpage]{acronym}
\usepackage{cleveref}

\begin{document}

\date{}

\title{\Large \bf MIRAGE: Multi-Binary Image Risk Assessment with Attack Graph Employment}

\author{}
\author{{\rm David~Tayouri, Telem~Nachum, Asaf~Shabtai}\\
\textit{Dept. of Software and Information Systems Engineering} \\
\textit{Ben-Gurion University of the Negev}\\
Beer-Sheva, Israel \\
\{davidtay,telemn\}@post.bgu.ac.il, shabtaia@bgu.ac.il
}

\maketitle

\begin{abstract}
Attackers can exploit known vulnerabilities to infiltrate a device's firmware and the communication between firmware binaries, in order to pass between them.
To improve cybersecurity, organizations must identify and mitigate the risks of the firmware they use.
An attack graph (AG) can be used to assess and visually display firmware's risks by organizing the identified vulnerabilities into attack paths composed of sequences of actions attackers may perform to compromise firmware images.
In this paper, we utilize AGs for firmware risk assessment.
We propose MIRAGE (Multi-binary Image Risk Assessment with Attack Graph Employment), a framework for identifying potential attack vectors and vulnerable interactions between firmware binaries; MIRAGE accomplishes this by generating AGs for firmware inter-binary communication.
The use cases of the proposed firmware AG generation framework include the identification of risky external interactions, supply chain risk assessment, and security analysis with digital twins.
To evaluate the MIRAGE framework, we collected a dataset of 703 firmware images.
We also propose a model for examining the risks of firmware binaries, demonstrate the model's implementation on the dataset of firmware images, and list the riskiest binaries.
\end{abstract}


\section{Introduction}
Device firmware, which is computer software that provides low-level control for a device’s hardware, may include many binaries to provide the device's functionality.
Complex tasks, such as communication (connecting to the Internet, managing packets, Bluetooth communication) and web interface management, require the running of multiple processes, i.e., firmware binaries, and these processes communicate, share data, and have dependencies; a description of firmware is provided in Section~\ref{subsec:Firmware}.

The firmware of embedded devices is commonly targeted in cyberattacks~\cite{michaelattack, sattlerattack}.
Every service that interacts with the external world (e.g., a device’s web service or a binary that parses communication packets) is considered a potential risk and part of the attack surface.
Malicious code injected in Internet-facing binaries may propagate between processes within the firmware.
\ac{IoT} devices' firmware, which may contain known vulnerabilities that hackers can easily find and exploit, pose a major security threat~\cite{nuspire2023ongoing}.
According to Bitdefender~\cite{bitdefender2023iot}, US and European homes respectively have an average of 46 and 25 devices connected to the Internet; and home networks see an average of eight attacks against devices every day; Section~\ref{subsec:FirmwareSec} provides a description of firmware security issues.

To improve cybersecurity, organizations must identify and mitigate the risks of the firmware they utilize.
For every possible threat, there may be several countermeasures; since it is infeasible to implement all countermeasures, organizations also need to assess their firmware's risks, prioritize them, and identify the security measures that will best reduce the threats to an acceptable level.

Various attack modeling techniques can be used to assess risks and visually present them.
The \ac{AG} is a popular method for visually representing cyber risks.
This risk assessment method presents attack states, transitions between them, and the related vulnerabilities as a graph~\cite{phillips1998graph,zeng2019survey}.
AGs organize identified vulnerabilities into attack paths composed of sequences of actions an attacker can take to compromise multi-binary software images. 
AGs can also help identify the attack paths most likely to succeed.
Consequently, their use enables security teams to prioritize the risks associated with firmware and determine which vulnerabilities to patch first.

AGs have two main advantages over other risk assessment methods.
First, an \ac{AG} models the interactions between vulnerabilities (multi-stage attacks) and the attacker's lateral movements (multi-host attacks). 
Second, an \ac{AG} is built for a specific target environment.

In this research, we employ the \ac{LAG}, a directed graph in which the leaves represent facts about the system, the internal nodes represent actions (attack steps) and their consequences (privileges), and the root represents an attacker’s final goal.
MulVAL, a well-known open-source framework for constructing LAGs~\cite{mulval}, is scalable, extensible, and commonly used by researchers; a description of the MulVAL framework is presented in Section~\ref{subsec:Mulval}.
AGs, and MulVAL in particular, were designed for and tested in standard enterprise or \ac{ICS} environments and are typically used in those environments. 
To the best of our knowledge, we are the first to use AGs in firmware risk assessment.

Usually, firmware contains many binaries that interact on different channels, such as direct calls, environment variables, named pipes, and shared memory.
Since firmware can run different types of software, each of which may have different vulnerabilities and interact differently with users and the external world, there may be many possible attack paths.

In this paper, we propose MIRAGE---Multi-binary Image Risk Assessment with Attack Graph Employment, a framework for identifying potential attack vectors and vulnerable interactions between firmware binaries; MIRAGE accomplishes this by generating AGs for firmware inter-binary communication.
The proposed MIRAGE framework's extendable \ac{AG} generation tool enables modeling new attack scenarios, thereby contributing to the detection and understanding of new risks.
The framework's use of static analysis enables identification of the potential attack surface, interactions between the binaries, and their vulnerabilities (see Section~\ref{sec:Mirage}).
MIRAGE first finds all of the binaries in the given firmware and builds the firmware's topology, i.e., the different ways binaries may interact.
It then identifies network-facing binaries (which constitute the attack surface) and \ac{OSS} binaries (which represent potential insider attackers that may have infiltrated the firmware through the supply chain).
Next, MIRAGE identifies all of the binaries' known vulnerabilities and generates an AG.
The use cases of the proposed firmware AG generation framework include the identification of risky external interactions, supply chain risk assessment, and security analysis with digital twins (see Section~\ref{subsec:UseCases}).

To evaluate the MIRAGE framework, we collected a dataset of 703 firmware images (see Section~\ref{subsec:Dataset}), attempted to generate an AG for each firmware image, and analyzed the results.
We found that 92\% of the firmware images we tested had an AG.
We also found that each firmware contained, on average, \~1.7 attack points, more than 3 potentially compromised instances of \ac{OSS}, and 2.5 vulnerable binaries, with a maximum of 13, 20, and 24, respectively.
To assess the firmware binaries' risk, in this paper, we also propose a model that examines the impact and likelihood of risks associated with each binary and list the binaries with the highest risks in our dataset; the full evaluation results are presented in Section~\ref{subsec:Analysis}.

The risk assessment of firmware is essential for improving the security of today’s widely used \ac{IoT} devices and the performance of the security teams responsible for defending them.
The main contributions of this paper are:
\begin{itemize}
    \item We propose an AG-based framework for firmware risk assessment.
    \item We evaluate the proposed framework by generating AGs for a dataset of 703 multi-binary firmware images.
    \item We propose a model for firmware binaries' risk assessment, demonstrate the model on our firmware image dataset, and list the riskiest binaries.
    \item To support open science and enable further research, we make our system available to the research community under an open-source license.
\end{itemize}

\section{Background} \label{sec:background}
\subsection{Firmware} \label{subsec:Firmware}
Firmware is defined as a specific class of computer software that provides low-level control for a device’s hardware.
The complexity of firmware depends directly on the functionality of the device it serves.
Devices designed for just one simple task may contain very basic firmware consisting of a few assembly instructions (e.g., emergency braking systems in cars and motion sensing lights).
However, the firmware of devices that are designed to handle multiple tasks (like most \ac{IoT} devices) should have a robust and efficient way of managing resources.
Vendors commonly use lightweight operating systems like embedded Linux or Android to operate multi-task devices; one large-scale experiment analyzed tens of thousands of firmware samples and found that 86\% of them were Linux-based~\cite{costin2014large}.

The firmware of complex devices may include a large number of binaries.
Tasks such as connecting to the Internet, managing a web interface, managing packets (routers), and communicating with Bluetooth devices require the running of multiple processes and system services, each of which originates from executable binaries and shell scripts from the firmware’s file system.
In multi-binary systems, processes communicate and share data using inter-process communication (IPC) mechanisms, and there are dependencies between processes.
Knowledge regarding the type of interaction and data transfer is critical for examining the nature of the relationship between firmware’s binaries and assessing the risks associated with firmware.

\subsection{Firmware Security Issues} \label{subsec:FirmwareSec}
Embedded device firmware has become one of the most common targets of cyberattacks~\cite{michaelattack, sattlerattack, assessingenterprise}.
Attacks on web services are among the most widely seen attacks in the wild due to their many known vulnerabilities.
\ac{IoT} devices with web services, such as wireless routers and web cameras, suffer more attacks than other devices~\cite{iotthreatreport, khandelwal2018thousands}.

The web services of a device are exposed to the Internet by design, and the ability of an attacker to send them direct HTTP requests makes those services an obvious target.
However, searching for vulnerabilities in Internet-facing processes is insufficient, since data commonly propagates between processes within firmware.
For example, string literals on web interfaces are commonly shared between front-end files and back-end binaries to encode user input~\cite{chen2021sharing, liu2022finding}.

Attackers' capabilities can expand when a device supports communication protocols like Bluetooth and Wi-Fi.
Any dedicated binary responsible for parsing packets of communication protocols is part of the attack surface.
Although it is relatively harder to find vulnerabilities in this kind of binary, every service that interacts with the outside world should be considered a potential risk to the device~\cite{lonzetta2018security, zubair2019exploiting}.

In addition, a firmware security issue may arise when several vendors depend on similar resources such as tools, \ac{SDKs}, and \ac{OSS} repositories.
Since different devices often share similar capabilities, vendors do not need to develop them in-house and instead rely on the capabilities of third parties.
In this case, devices may be branded differently but actually run the same or similar binaries.
When it comes to vulnerabilities, sharing vulnerable code may lead to multiple compromised devices from different brands.
When developing new firmware, vendors should consider the risk when determining which functionalities will be developed in-house and which will be provided by third parties.

Furthermore, legacy code and libraries can be misused by an attacker.
Sometimes, firmware contains old source code that has already been discovered as a vulnerable component.
Vendors' developers mistakenly might use working but vulnerable versions of libraries, binaries, and even kernels in the firmware.
Once hostile parties are aware that the device uses known vulnerable libraries, they can exploit it using a known exploit (e.g., published in Exploit DB~\cite{exploitdb}), leading to a compromised system.

Firmware can also be stealthily compromised.
US authorities published such a case in a joint cybersecurity advisory in September 2023~\cite{cisa2023prc}.
In the advisory, they described the activity of the People’s Republic of China (PRC)-linked cyber actor known as BlackTech, which has demonstrated capabilities in modifying router firmware without detection and exploiting routers’ domain-trust relationships to pivot from international subsidiaries to their primary targets.
BlackTech members use custom malware, dual-use tools, and living off the land tactics, such as disabling logging on routers, to conceal their operations in order to target the government, industrial, technology, media, electronics, and telecommunication sectors.

\subsection{Attack Graphs and MulVAL} \label{subsec:Mulval}
An AG is a model that enables researchers and security practitioners to provide a visual representation of events that can lead to a successful attack scenario.
Various types of AGs have been proposed.
Hong et al.~\cite{hong2017survey} performed a survey reviewing all of the modeling techniques and AG generation tools presented in the literature.
The most common AG representations include the attack tree (AT), state graph (SG), exploit dependency graph (EDG), logical attack graph (LAG), and multiple prerequisite attack graph (MPAG) representations.

In this research, we utilize MulVAL (multi-host, multi-stage vulnerability analysis language), an open-source publicly available logic-based attack graph generation tool~\cite{ou2005logic}. 
MulVAL is based on the Datalog modeling language, a subset of the Prolog logic programming language. 
In MulVAL, Datalog is used to represent two types of entities:
\begin{itemize}
\item \textit{Facts}: network topology and configuration, security policies, and known vulnerabilities.
\item \textit{Rules} (also known as interaction rules): the interactions between components in the network.
\end{itemize}

Facts and rules are defined by applying a predicate $p$ to some arguments: $p(t_1,...,t_k)$.
Each $t_i$ can be either a constant or a variable.
The Datalog syntax indicates that a constant is an identifier that starts with a lowercase letter, while a variable begins with an uppercase letter.
A wildcard expression can be defined by the underscore character ('$\_$').
A sentence in MulVAL is defined as a Horn clause of literals:
$$L_0 :- L_1,...,L_n$$ \label{def:IR}
Where $L_0$ is defined as the head, and $L_1,...,L_n$ comprise the body of the sentence.
Each $L_i$ in the body can be either a fact or an interaction rule.
Body $(L_1,...,L_n)$ literals are preconditions for the head ($L_0$): if the body literals are true, then the head literal is also true.
A sentence with an empty body is called a fact.
For example, the following fact states that there is an identified vulnerability \texttt{CVE-2002-0392} in the \texttt{httpd} service running on the \texttt{webServer01} instance:
\begin{lstlisting}[basicstyle=\ttfamily\scriptsize,language=Python]
vulExists(webServer01, "CVE-2002-0392", httpd).
\end{lstlisting}
A sentence with a nonempty body is called a rule.
For example, the rule in Listing~\ref{lst:fileprot} says that if a \texttt{User} has ownership of \texttt{Path} on \texttt{Host}, and if an owner of \texttt{Path} on \texttt{Host} has the specified \texttt{Access}, then the \texttt{User} on \texttt{Host} can have the specified \texttt{Access} to \texttt{Path}.
\begin{lstlisting}[basicstyle=\ttfamily\scriptsize,language=Python,label=lst:fileprot, caption=Interaction rule example]
localFileProtection(Host, User, Access, Path) :-
    fileOwner(Host, Path, User),
    ownerAccessible(Host, Access, Path).
\end{lstlisting}

MulVAL's reasoning engine estimates the effect of the identified vulnerabilities on the system.
This estimation is performed by applying the defined set of interaction rules on the generated facts.

As a \ac{LAG}, MulVAL rules can be extended to represent known \ac{TTPs}.
Being generic and extensible, LAGs support a large class of threat models characterized by attackers' goals, capabilities, and resources.  
LAGs can also support the various levels of attacker capabilities if they are defined as preconditions for exploits~\cite{malzahn2020automated, wang2020cvss}. 

In 2013, Yi et al.~\cite{yi2013overview} compared several academic and commercial attack graph generation tools (Topological Vulnerability Analysis, Attack Graph Toolkit, NetSPA, MulVAL, Cauldron, FireMon, and Skybox View).
The authors concluded that MulVAL is the most extendable and scalable framework. While commercial tools may be more scalable and user-friendly, they are not open-source and are thus less suitable for academic research.

MulVAL benefits from the advantage of extensibility: its underlying reasoning engine is written in a logical programming language, which enables users to extend functionality by writing custom rules.
We leveraged this capability and defined some new \ac{IRs} in order to generate AGs for firmware.

\section{Proposed Method} \label{sec:Mirage}
To assess the risk of firmware, we developed MIRAGE (Multi-binary Image Risk Assessment with Attack Graph Employment), which uses a LAG generation tool.
The MIRAGE framework's code is available at~\cite{miragegithub}.

\begin{figure}[ht]
    \centering
    \includegraphics[width=0.95\columnwidth]{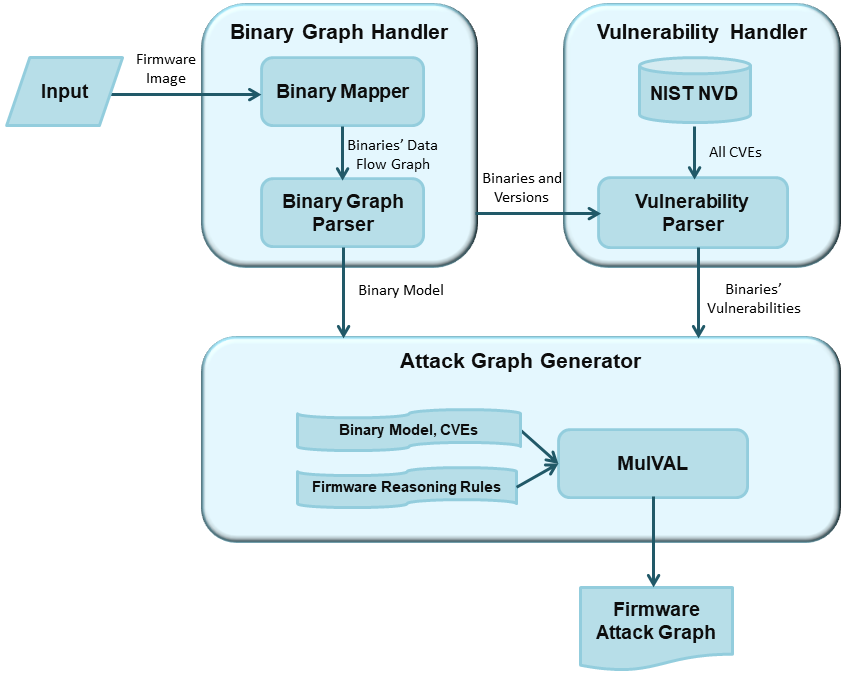}
    \caption{MIRAGE architecture.}
    \label{fig:MirageArchitecture}
\end{figure}

Figure~\ref{fig:MirageArchitecture} presents the MIRAGE framework's architecture.
The framework consists of three modules, which are described below: the Binary Graph Handler generates the firmware binary model (see Section~\ref{BinaryGraphHandler}); the Vulnerability Handler generates the list of vulnerabilities found in the given firmware (see Section~\ref{VulnerabilityHandler}); and the Attack Graph Generator runs MulVAL to generate an AG (see Section~\ref{AttackGraphGenerator}).
Section~\ref{subsec:UseCases} presents several use cases for the multi-binary firmware AG generation framework.

\subsection{Binary Graph Handler} \label{BinaryGraphHandler}
In this module, the Binary Mapper scans the firmware image and finds the binaries that interact with the outside world (Internet or users) and the connections between the binaries in the firmware.
As input, the Binary Mapper receives a compressed firmware image.
The Binary Mapper's first step is to decompress the firmware image and extract its file system, using Binwalk~\cite{binwalk}.
After extracting the firmware’s file system, the Binary Mapper scans for any executable binary that exists in the file system, using the Linux \texttt{file} capability to determine if a file is an executable.
The list of all executable binaries in the firmware is fundamental to AG generation, as each executable is a node in the image topology graph.
The Binary Mapper's next two steps are finding network-facing binaries and the connections between different binaries.
This is achieved by employing two Karonte~\cite{redini2020karonte} components: Border Binaries Discovery and  Binary Dependency Graph.

To find Internet-facing binaries, we use Katonte's Border Binaries Discovery component, which leverages the observation that network binaries are firmware elements that receive and parse network data.
Therefore, the component identifies binaries that read from network sockets.
Then we use Karonte’s Binary Dependency Graph component, which detects data dependencies among the binaries.
It does so by checking if binaries interact using environment variables, sockets, or files.

We included a new method of detecting inter-process interaction between binaries -- checking if a binary runs another binary.
This is done by searching \texttt{execv} and \texttt{system} calls followed by a binary name.

This step results in a binary data graph that contains all the executable binaries of the firmware, the connection between the binaries, and the data flow among connected binaries.
The Binary Mapper's output is the list of the binaries and the connections between them in JSON format.
An example of a binary data flow graph is provided in Listing~\ref{lst:BinaryGraphSample}.
\begin{lstlisting}[basicstyle=\ttfamily\scriptsize,language=Python,label=lst:BinaryGraphSample, caption=Example of binary data flow graph]
"fW_name": "NETGEAR_R7800_da9",
"graph": {
  ...
  "uhttpd": {
    "peers": [
      {
        "name": "opkg",
        "type": "environment",
        "info": ["PATH"]
      },
      {
        "name": "proccgi",
        "type": "environment",
        "info": ["QUERY_STRING","PATH_INFO",...]
      },
      {
        "name": "net-cgi",
        "type": "exec",
        "info": ""
      },
      {
        "name": "INTERNET",
        "type": "border_binary",
        "info": ""
      },
      ...
    ],
    "version": []  
  },
  "opkg": {
    "peers": [
      {
        "name": "afpd",
        "type": "exec",
        "info": ""
      },
      {
        "name": "wget",
        "type": "environment",
        "info": ["ftp_proxy", "http_proxy"]
      },
      {
        "name": "busybox",
        "type": "environment",
        "info": ["PATH"]
      },
      {
        "name": "tar",
        "type": "exec",
        "info": ""
      },
      ...
    ],
    "version": ["0.1.8", "v1", "v2"]
  },
}
\end{lstlisting}

\begin{figure}[ht]
    \centering
    \includegraphics[width=0.95\columnwidth]{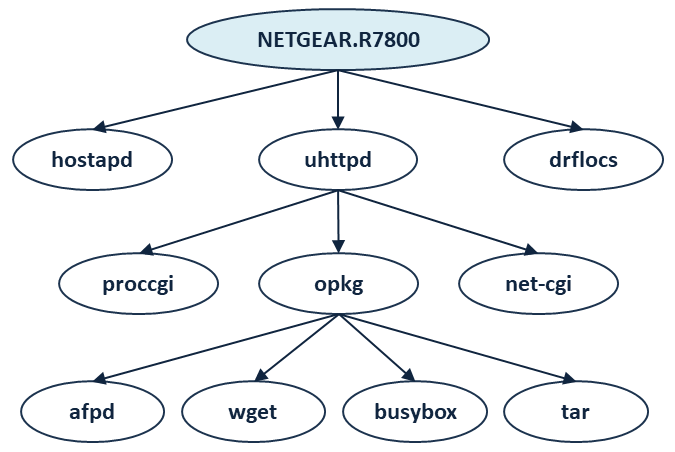}
    \caption{Sample firmware graph.}
    \label{fig:SampleFirmwareGraph}
\end{figure}

Figure~\ref{fig:SampleFirmwareGraph} visually presents a partial binary data flow graph of Listing~\ref{lst:BinaryGraphSample}.
The results of Binary Mapper are passed to the Binary Graph Parser to create MulVAL facts (binary model).
\texttt{externalInteraction} represents Internet-facing binaries, and \texttt{dataFlow} represents interactions between binaries.
An example of a binary model (the facts generated for Listing~\ref{lst:BinaryGraphSample}) is provided in Listing~\ref{lst:TopologyFactSample}.
\begin{lstlisting}[basicstyle=\ttfamily\footnotesize,language=Python,label=lst:TopologyFactSample, caption=Example of binary model]
dataFlow(uhttpd, opkg, 'environment').
dataFlow(uhttpd, proccgi, 'environment').
dataFlow(uhttpd, net_cgi, 'exec').
dataFlow(opkg, afpd, 'exec').
dataFlow(opkg, wget, 'environment').
dataFlow(opkg, busybox, 'environment').
dataFlow(opkg, tar, 'exec').
externalInteraction('Internet', uhttpd, Internet).
bugHyp(wget, 'LOCAL', 'Undefined').
\end{lstlisting}

Another output of the Binary Mapper is the list of binaries and their versions, a list which serves as input for the Vulnerability Handler module (described in Section~\ref{VulnerabilityHandler}).
We use two main methods for determining the binaries' versions. First, we try to extract the version string dynamically.
For binaries that can handle user input arguments, it is very common to have a built-in functionality of printing the version to stdout.
Running the binary with flags such as ‘-v,’ ‘--v,’ ‘-version,’ and even the help command may print a binary’s version in the help message.
To leverage this behavior, we use the QEMU tool, a generic and \ac{OSS} machine emulator and virtualizer~\cite{qemu}.
We run the binary with several version-related arguments that may print the version to stdout.

The second method of extracting binaries' versions is to use the version scanner component from Intel’s OSS tool - the CVE Binary Tool~\cite{cvebintool}.
The version scanner component supports version extraction of over 200 common binaries and, in some cases, improves our module's ability to extract the binaries' versions.
In some cases, the module fails to extract the version of a binary.
Binaries that do not have the functionality of printing the version dynamically and for which Intel’s CVE Binary Tool fails in extracting the binaries are left without the version noted.
An example of a list of a firmware's binaries and versions is provided in Listing~\ref{lst:FirmwareBinariesSample}.

\begin{lstlisting}[basicstyle=\ttfamily\footnotesize,language=Python,label=lst:FirmwareBinariesSample, caption=Example of list of firmware's binaries and versions]
busybox, 1.4.2
tar, 1.23
tar, 4.2BSD
hostapd, 2.5
uhttpd, *
opkg, 0.1.8
opkg, 1
proccgi, *
net-cgi, *
drflocs, 1.15.10
afpd, *
wget, 1.13.4
\end{lstlisting}

\subsection{Vulnerability Handler} \label{VulnerabilityHandler}
As input, The Vulnerability Handler receives the list of binaries and versions from the Binary Graph Handler module.
The Vulnerability Handler's first step is to filter the entire NIST \ac{NVD} \ac{CVE} list by the received list of binaries.
The result is a list of vulnerabilities relevant to the given firmware.
The Vulnerability Parser performs the next step of creating MulVAL facts for the list of filtered vulnerabilities.
\texttt{vulExists} is used to present binaries' vulnerabilities.
An example of the binary vulnerability facts generated is provided in Listing~\ref{lst:VulnerabilityFactSample}.
\begin{lstlisting}[basicstyle=\ttfamily\footnotesize,language=Python,label=lst:VulnerabilityFactSample, caption=Example of binary vulnerability facts]
vulExists('CVE-2017-13089', wget, 'NETWORK',
  'availability_loss', 'HIGH').
vulExists('CVE-2018-1000517', busybox, 'NETWORK',
  'availability_loss', 'HIGH').
vulExists('CVE-2021-38511', tar, 'NETWORK',
  'data_modification', 'MEDIUM').
vulExists('CVE-2022-23303', hostapd, 'NETWORK',
  'availability_loss', 'MEDIUM').
\end{lstlisting}

\subsection{Attack Graph Generator} \label{AttackGraphGenerator}

To generate an AG for the given image, we run MulVAL using the input facts created for the binary model and CVEs.
The facts are presented by one of the predicates listed in Listing~\ref{lst:BasicFacts}.
\texttt{externalInteraction} defines the network-facing binaries and binaries interacting with users.
To define interactions between the binaries, \texttt{dataFlow} is used.
\texttt{vulExists} defines the binaries' vulnerabilities.

\begin{lstlisting}[basicstyle=\ttfamily\scriptsize,language=Python,label=lst:BasicFacts, caption=Firmware facts]
externalInteraction(Source, Software, InteractionType).
dataFlow(Software1, Software2, FlowType).
vulExists(CveId, SW, AccessVector, LoseTypes, Severity).
\end{lstlisting}

In addition to facts, MulVAL requires firmware IRs.
The reasoning rules we define for firmware AG generation are presented in Listing~\ref{lst:InteractionRules}.
The first and second rules say that software (i.e., a binary) is vulnerable if it can be accessed from the Internet and has a known CVE (critical and high severity, respectively) that can be exploited remotely.
The third and fourth rules say that if SW1 is vulnerable, there is a flow between SW1 and SW2, and SW2 has a known CVE that can be exploited remotely (critical and high severity, respectively), then SW2 is vulnerable.

\begin{lstlisting}[basicstyle=\ttfamily\scriptsize,language=Python,label=lst:InteractionRules, caption=Firmware interaction rules]
vulnerableSoftware(SW) :-
  externalInteraction('Internet', SW, _interactionType),
  vulExists(_cveId, SW, 'NETWORK', _lose_types, 'CRITICAL').

vulnerableSoftware(SW) :-
  externalInteraction('Internet', SW, _interactionType),
  vulExists(_cveId, SW, 'NETWORK', _lose_types, 'HIGH').

vulnerableSoftware(SW2) :-
  vulnerableSoftware(SW1),
  dataFlow(SW1, SW2, _flowType),
  vulExists(_cveId, SW2, 'NETWORK', _loseTypes, 'CRITICAL')
    
vulnerableSoftware(SW2) :-
  vulnerableSoftware(SW1),
  dataFlow(SW1, SW2, _flowType),
  vulExists(_cveId, SW2, 'NETWORK', _loseTypes, 'HIGH')
    
\end{lstlisting}

\begin{figure*}[ht]
    \centering
    \begin{subfigure}[b]{\columnwidth}
    \centering
    \includegraphics[width=0.95\columnwidth]{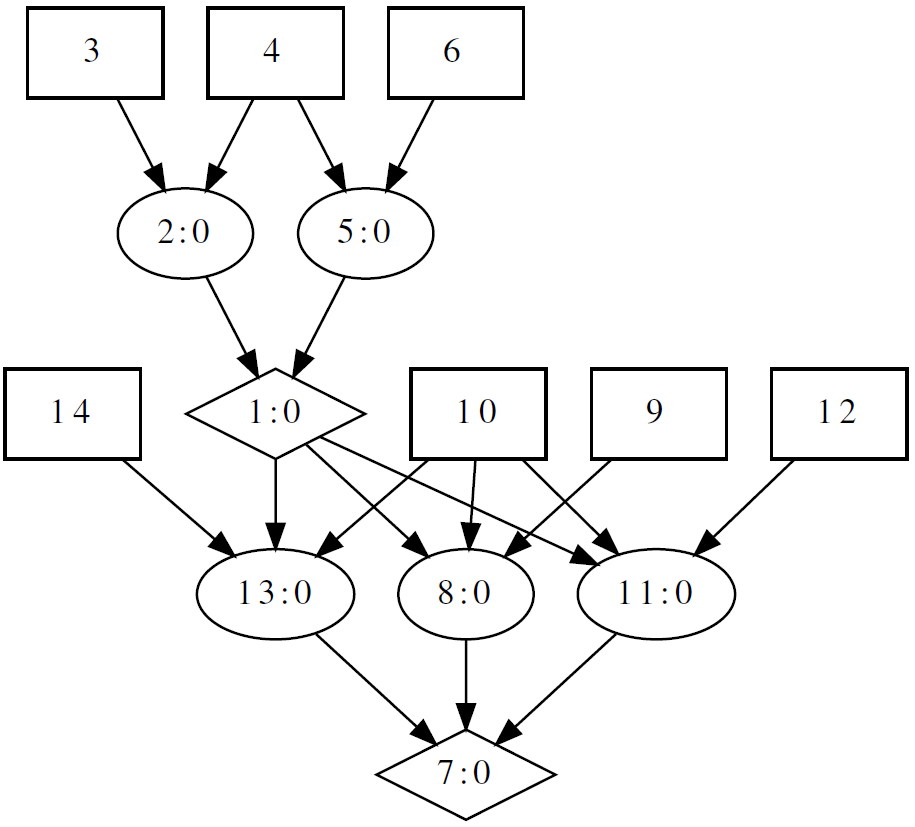}
    \caption{Partial AG of NETGEAR AC1450.}
    \label{fig:SampleAG-External-A}
    \end{subfigure}
    \centering
    \begin{subfigure}[b]{\textwidth}
    \centering
    \includegraphics[width=0.95\columnwidth]{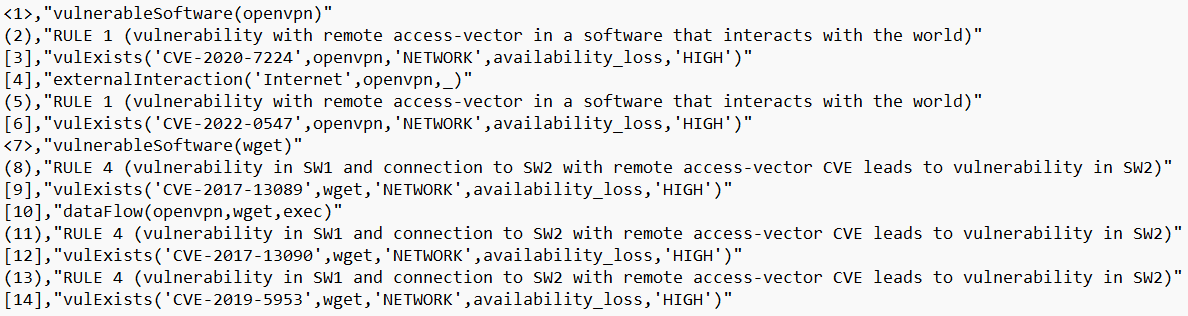}
    \caption{AG node interpretation.}
    \label{fig:SampleAG-External-B}
    \end{subfigure}
    \caption{Partial AG for an external threat in a sample firmware.}
\end{figure*}

Figures~\ref{fig:SampleAG-External-A} and~\ref{fig:SampleAG-External-B} present a partial AG for an external threat in a sample firmware.
We can see that there is an Internet-facing binary \texttt{openvpn} with two CVEs with a remote access vector (the attack surface of the image).
This binary interacts with \texttt{wget}, which has several high-severity CVEs that can be exploited remotely.
The AG shows that \texttt{openvpn} is vulnerable, and it also makes \texttt{wget} vulnerable (an attack vector).

As mentioned in Section~\ref{subsec:FirmwareSec}, firmware can be stealthily compromised; therefore, to comprehensively assess firmware risks, we should also assume an internal threat, i.e., third-party binaries are already infected.
To achieve this, we should identify the attack paths that can be caused by compromised third-party binaries.
We start by finding the OSS binaries in the firmware.
To do this, we intersect the firmware binary list with the NIST NVD CVE binaries.
If a binary (regardless of its version), ever had a known CVE, the binary is considered OSS.
Then, based on our assumption (hypothesis) that the OSS binaries have undocumented vulnerabilities, we use the predicate \texttt{bugHyp}, presented in Listing~\ref{lst:BugHypIRs}, which enables the definition of potentially compromised OSS binaries.
To generate AGs based on the hypothesis of internal threat, we define the IRs presented in Listing~\ref{lst:BugHypIRs}.
The first IR in the listing defines software as compromised if it has a potential vulnerability (bug).
The second rule says that if SW1 is potentially vulnerable, there is a flow between SW1 and SW2, and SW2 has a known critical CVE that can be exploited remotely, then SW2 is vulnerable.
The third rule is the same as the previous one, except in this case, SW2 must have a CVE with high severity.

\begin{lstlisting}[basicstyle=\ttfamily\scriptsize,language=Python,label=lst:BugHypIRs, caption=Potentially compromised binary predicates]
bugHyp(Software, AccessVector, LoseTypes).

potentiallyVulnerableSoftware(Software) :-
  bugHyp(Software, _accessVector, _loseTypes)

vulnerableSoftware(SW2) :-
  potentiallyVulnerableSoftware(SW1),
  dataFlow(SW1, SW2, _flowType),
  vulExists(_cveId, SW2, 'NETWORK', _loseTypes, 'CRITICAL')

vulnerableSoftware(SW2) :-
    potentiallyVulnerableSoftware(SW1),
    dataFlow(SW1, SW2, _flowType),
    vulExists(_cveId, SW2, 'NETWORK', _loseTypes, 'HIGH')
    
\end{lstlisting}

\begin{figure*}[h!t]
    \centering
    \begin{subfigure}[b]{\textwidth}
    \centering
    \includegraphics[width=0.6\textwidth]{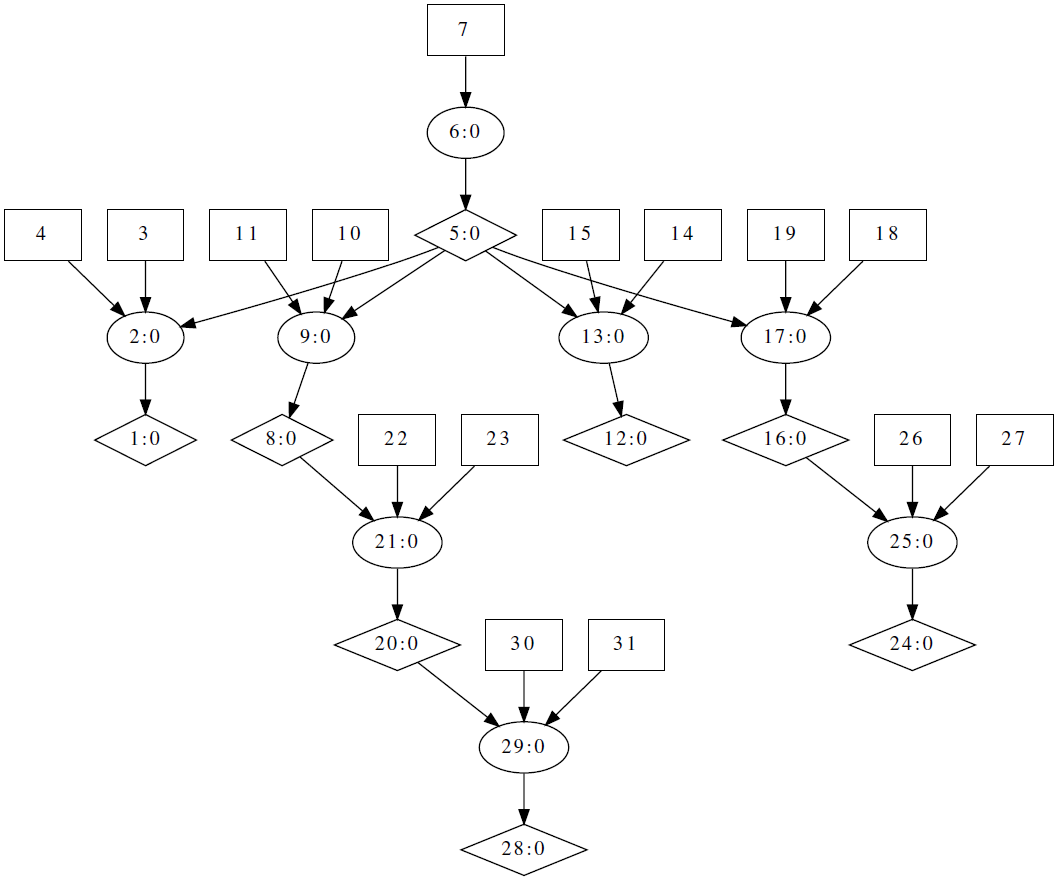}
    \caption{Partial AG of NETGEAR AC1450.}
    \label{fig:SampleAG-Hyp-A}
    \end{subfigure}
    \centering
    \begin{subfigure}[b]{\textwidth}
    \centering
    \includegraphics[width=0.9\textwidth]{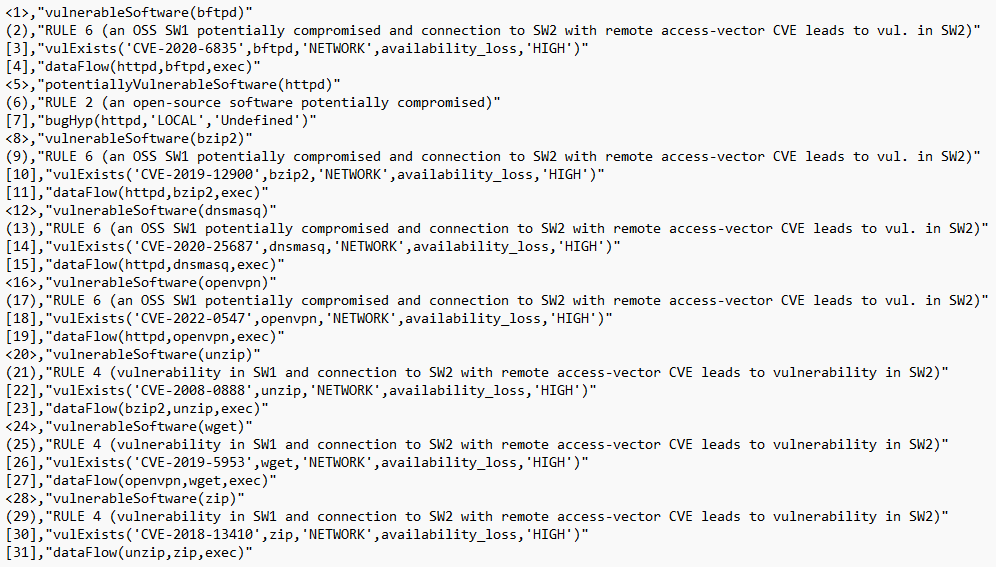}
    \caption{AG node interpretation.}
    \label{fig:SampleAG-Hyp-B}
    \end{subfigure}
    \caption{Partial AG for an internal threat in a sample firmware.}
\end{figure*}

Figures~\ref{fig:SampleAG-Hyp-A} and~\ref{fig:SampleAG-Hyp-B} present a partial AG for an internal threat in a sample firmware.
We can see that \texttt{httpd} is a potentially exploited OSS binary (supply chain threat - an attacker has compromised the image through an OSS binary).
This binary interacts with other binaries that have high-severity CVEs that can be exploited remotely, \texttt{bftpd}, \texttt{bzip2}, \texttt{dnsmasq}, and \texttt{openvpn}.
\texttt{openvpn} interacts with \texttt{wget}.
\texttt{bzip2} interacts with \texttt{unzip}, which in turn interacts with \texttt{zip}.
The AG shows how one potentially compromised binary (\texttt{httpd}) can create attack vectors through another six binaries with vulnerabilities, leading to \texttt{wget} and \texttt{zip} and making them accessible to an attacker.

\begin{figure}[ht]
    \centering
    \includegraphics[width=0.95\columnwidth]{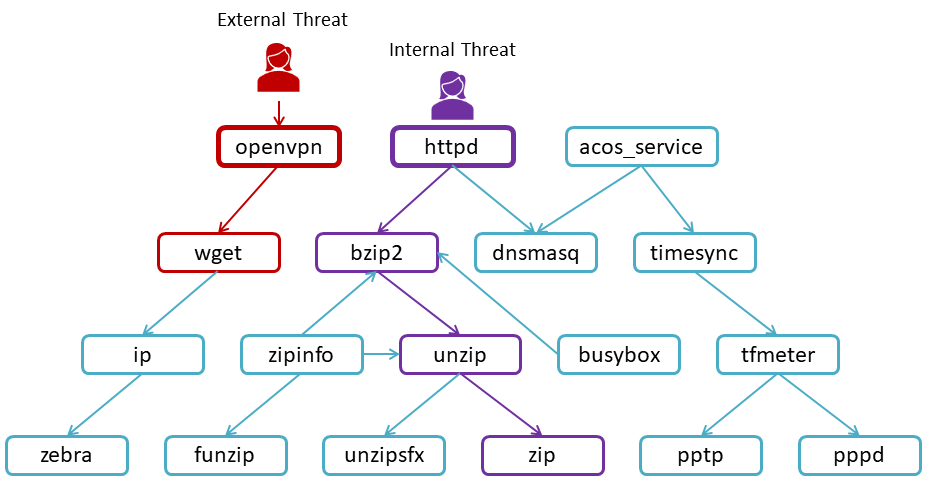}
    \caption{Partial topology of a sample firmware with highlighted attack paths.}
    \label{fig:SampleFirmwareTopologyWithAG}
\end{figure}

Figure~\ref{fig:SampleFirmwareTopologyWithAG} presents a partial topology of NETGEAR AC1450 with two highlighted attack paths - one for an external threat (red) and one for an internal threat (purple).

\subsection{Multi-Binary AG Use Cases} \label{subsec:UseCases}
The generation of AGs for multi-binary images enables several use cases:
\begin{itemize}
\item External interactions discovery: identifying external interactions, e.g., remote admin connection or user input, which can serve as the attack surface from which an adversary can attack a software system or firmware in a specific environment.
This can help in identifying vulnerable entry points and strengthening them (e.g., by patching) or closing them.
\item Supply chain risk assessment: determining if (and which) third-party components are part of potential attack vectors; determining what an attacker can do if there is a library with vulnerabilities.
This can help in deciding whether to use the library.
\item Security analysis with digital twins: applying generated AGs to digital twins for security analysis.
For example, Marksteiner et al.~\cite{marksteiner2021using} suggested a method in which automotive firmware is transformed into a cyber digital twin and continuously evaluated for vulnerability detection.
\end{itemize}

\section{Evaluation} \label{Evaluation}
\subsection{Dataset} \label{subsec:Dataset}
To evaluate the proposed framework, we collected a dataset of 703 firmware images from Karonte~\cite{redini2020karonte}.
This dataset enabled a comprehensive quantitative analysis of our framework.
The firmware vendors include Airlink101, Belkin, Buffalo, Cisco, D-Link, forceWare, Foscam, Huawei, Inmarsat, Linksys, NETGEAR, NVIDIA, OpenWrt, Poly, Qualcomm, Supermicro, Synology, TP-Link, TRENDnet, Tenda, Tomato, Ubiquiti, Verizon, and Zyxel.
We analyzed firmware whose architecture is aligned with the capabilities of angr, a binary analysis framework (used by Karonte) that integrates many binary analysis techniques~\cite{shoshitaishvili2016sok}.
In particular, this tool supports ARM, AArch64, and PowerPC architectures.

For each firmware, we run the MIRAGE modules as follows: the Binary Mapper is used to identify the firmware's binary topology (Internet-facing binaries and connections between the binaries); the Binary Graph Parser is used to generate a binary model; the Vulnerability Handler module is used to identify the binaries' versions and their vulnerabilities; and the AG Generator module is used to generate AGs for the firmware.

\subsection{Analysis} \label{subsec:Analysis}
After generating AGs for all of the dataset's firmware, we analyzed the results.
Fifty-seven of the 703 firmware images did not have an AG, i.e., did not have a vulnerable attack surface or potentially compromised OSS binaries; this means that 92\% of the firmware images in our dataset had AGs.
For each AG, we performed the following steps.
First, we determined the number of attack points (the entry points an attacker can exploit).
This was done by counting the number of \texttt{externalInteraction} IRs in the firmware’s AG.
Then, we determined the number of potentially compromised \ac{OSS} binaries.
This was achieved by counting the number of \texttt{bugHyp} IRs in the firmware’s AG.
Finally, we determined the number of vulnerable software binaries, either software that is accessible from the external world (external attacker) or potentially compromised OSS (internal attacker).
This was done by counting the number of \texttt{vulnerableSoftware} and \texttt{potentiallyVulnerableSoftware} IRs in the firmware's AG.
Since this included the binaries counted in the previous two steps, we subtracted them from the total in order to identify any “new” assets an attacker could access.

For each step, the mean and max number are presented in Table~\ref{table:FirmwareBasicAnalysis}.
We can see that on average there are almost two attack points and more than three OSS potentially compromised in each firmware, with a maximum of 8 and 18, respectively.
In addition, there are, on average, 3.4 vulnerable binaries in each firmware, with a maximum of 24.

\begin{table}[htb]
    \centering
    \caption{Basic analysis of firmware}
    \small
    \begin{tabular}{| c | c | c |} 
    \hline
    \textbf{Subject} & \textbf{Mean} & \textbf{Max} \\ 
    \hline
    \# Attack Points & 1.7 & 13 \\
    \hline
    \# Potentially Compromised OSS & 3.1 & 20 \\
    \hline
    \# Vulnerable Binaries$^{\mathrm{(*)}}$ & 2.5 & 24 \\
    \hline
    \multicolumn{3}{l}{\footnotesize$^{\mathrm{(*)}}$Without attack points and potentially compromised OSS }\\
    \end{tabular}
    \label{table:FirmwareBasicAnalysis}
\end{table}

To understand the nature of the vulnerable binaries and assess the firmware binaries' risk, we propose a model that examines the impact and likelihood of risks associated with each binary.
First, we identified the vulnerable binaries that appear most frequently in AGs.
Next, we calculated each binary's mean number of interactions in all of the firmware's AGs.
This was done by counting the number of \texttt{dataFlow} IRs in the firmware’s AGs.
We consider the number of interactions per appearance of a binary as the binary's impact.
Then, we estimated the likelihood of each binary being exploited.
To do this, we used the \texttt{vulExists} IRs in the firmware’s AGs to find the binaries' CVEs, and for each CVE we searched the \ac{CISA} database~\cite{cisaexploitedcves} to determine whether the CVE has an exploit; if it does not have an exploit, we queried its \ac{EPSS}~\cite{firstepss}.
The EPSS is a community-driven effort to collect and analyze descriptive information about CVEs and estimate the likelihood that a vulnerability will be exploited.
The EPSS model produces a probability score between zero and one; the higher the score, the greater the probability that a vulnerability will be exploited.
For each binary, we stored the highest exploitability score (if a CVE has an exploit, the score is 100\%).
The product of the impact and the likelihood values represents a binary's risk.
We identified 142 vulnerable binaries in our dataset of 703 firmware images.
Table~\ref{table:RiskiestBinaries} presents the data collected for the 25 binaries with the greatest risk.

\begin{table*}[htb]
    \centering
    \small
    \caption{Binaries with the greatest risk (impact*likelihood)}
    \begin{tabular}{| c | c | c | c | c | c | c |} 
    \hline
    \textbf{Binary Name} & \textbf{Occurrences} & \textbf{Interactions} & \textbf{Impact} & \textbf{\# CVEs} & \textbf{Likelihood (\%)} & \textbf{Risk} \\
    \hline
    cups & 4 & 22 & 5.5 & 18 & 94.1 & 518 \\
    \hline
    lighttpd & 87 & 368 & 4.2 & 21 & 96.9 & 410 \\
    \hline
    openssl & 147 & 452 & 3 & 58 & 100 & 307 \\
    \hline
    php & 49 & 135 & 2.8 & 45 & 100 & 276 \\
    \hline
    mongoose & 8 & 108 & 13.5 & 9 & 18 & 244 \\
    \hline
    wget & 98 & 145 & 1.5 & 8 & 95.8 & 142 \\
    \hline
    tcpdump & 36 & 49 & 1.4 & 166 & 95 & 129 \\
    \hline
    dnsmasq & 238 & 617 & 2.6 & 14 & 46.2 & 120 \\
    \hline
    telnet & 24 & 30 & 1.3 & 2 & 92.6 & 116 \\
    \hline
    nginx & 54 & 64 & 1.2 & 8 & 95.8 & 114 \\
    \hline
    mysql & 19 & 22 & 1.2 & 19 & 97.3 & 113 \\
    \hline
    tftp\_hpa & 18 & 223 & 12.4 & 1 & 8.6 & 106 \\
    \hline
    memcached & 19 & 22 & 1.2 & 3 & 88.3 & 102 \\
    \hline
    ntp & 5 & 5 & 1 & 7 & 96.6 & 97 \\
    \hline
    squid & 1 & 1 & 1 & 49 & 96.7 & 97 \\
    \hline
    samba & 3 & 3 & 1 & 5 & 94.9 & 95 \\
    \hline
    vsftpd & 55 & 55 & 1 & 1 & 87.2 & 87 \\
    \hline
    rpcbind & 13 & 21 & 1.6 & 1 & 53 & 86 \\
    \hline
    ffmpeg & 24 & 32 & 1.3 & 39 & 60.4 & 80 \\
    \hline
    file & 19 & 233 & 12.3 & 3 & 6 & 74 \\
    \hline
    curl & 49 & 263 & 5.4 & 49 & 13 & 70 \\
    \hline
    asterisk & 3 & 1 & 0.3 & 30 & 96 & 32 \\
    \hline
    pg & 30 & 195 & 6.5 & 1 & 4.6 & 30 \\
    \hline
    traceroute & 4 & 11 & 2.8 & 1 & 11 & 30 \\
    \hline
    dhcpcd & 39 & 123 & 3.2 & 7 & 8.7 & 27 \\
    \hline
    \end{tabular}
    \label{table:RiskiestBinaries}
\end{table*}

We can see that the four riskiest OSS binaries are \texttt{cups}, \texttt{lighttpd}, \texttt{openssl}, and \texttt{php}.
In addition to being used for risk assessment, this information enables security practitioners to take security into account when examining the use of OSS binaries.
This information can also help them recommend or discourage the use of alternative binaries (in cases in which such alternatives exist).

\begin{figure*}[h!t]
    \centering
    \begin{subfigure}[b]{\textwidth}
    \centering
    \includegraphics[width=0.7\textwidth]{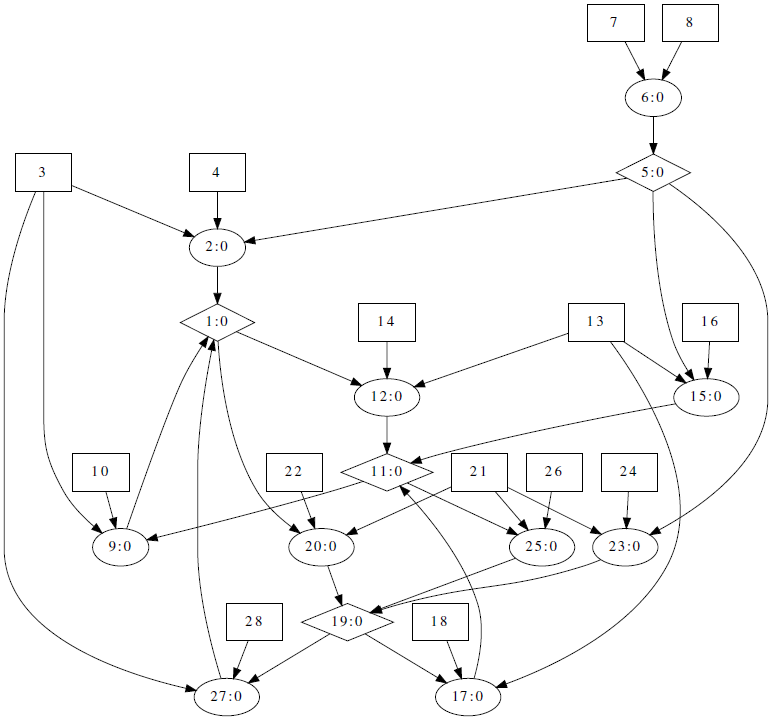}
    \caption{Partial AG for a sample firmware.}
    \label{fig:SampleAG2-A}
    \end{subfigure}
    \centering
    \begin{subfigure}[b]{\columnwidth}
    \centering
    \includegraphics[width=0.7\columnwidth]{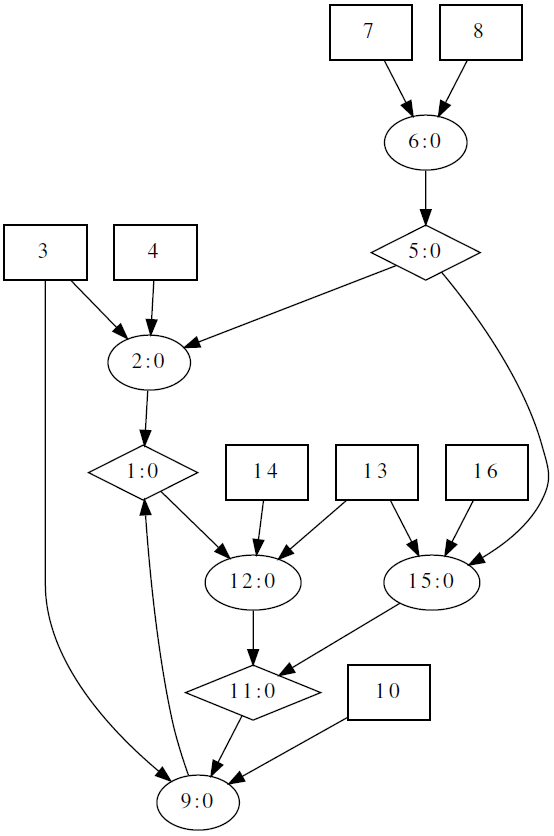}
    \caption{The AG after patching dhcpcd.}
    \label{fig:SampleAG2-B}
    \end{subfigure}
    \begin{subfigure}[b]{\columnwidth}
    \centering
    \includegraphics[width=0.7\columnwidth]{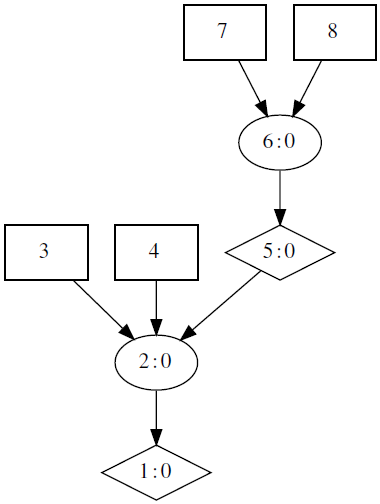}
    \caption{The AG after patching dhcpcd and cups.}
    \label{fig:SampleAG2-C}
    \end{subfigure}
    \caption{Partial AG, before and after patching binaries, for a sample firmware.}
\end{figure*}

The information presented in Table~\ref{table:RiskiestBinaries} can also facilitate the consideration of what-if questions, e.g., what will happen to an AG if we patch a risky binary.
Figure~\ref{fig:SampleAG2-A} presents a partial AG for an external threat in a sample firmware.
After patching the risky binary, \texttt{dhcpcd}, the AG shrinks, as seen in Figure~\ref{fig:SampleAG2-B}.
If we also patch \texttt{cups}, there is a single attack path, as seen in Figure~\ref{fig:SampleAG2-C}.

\section{Related Work} \label{sec:RelatedWork}
Previous works on firmware cybersecurity mainly focused on the following domains: fuzzing, \ac{SAST}, \ac{DAST}, formal validation, and risk assessment.

\subsection{Fuzzing}
Fuzzing is a well-known method for vulnerability detection in software systems, in which random input is provided to a program and the exceptions, such as crashes, are monitored.
Xie et al.~\cite{xie2017vulnerability} utilized static analysis and fuzzing to detect logical flaws in firmware.
Using their proposed method, they were able to detect two types of authentication bypass flaws.
A gray-box fuzzing scheme for Linux-based firmware was proposed by Zheng et al.~\cite{zheng2019efficient}.
The authors performed binary static analysis to generate useful inputs for efficient fuzzing.
Addressing the challenges faced in prior work, Yin et al.~\cite{yin2023rsfuzzer} proposed RSFuzzer, a hybrid gray-box fuzzing technique that can learn the input interface and format information and detect deeply hidden vulnerabilities triggered by invoking multiple system management interrupt handlers.
Gui et al.~\cite{gui2020firmcorn} presented FIRMCORN, a vulnerability-oriented fuzzer for IoT firmware; for efficient vulnerability mining, a vulnerable-code search algorithm was designed to obtain the fuzzing entry points based on the firmware's characteristics.
To detect memory corruption in lightweight IoT device firmware, Zhu et al.~\cite{zhu2019fiot} introduced FIoT, in which a fuzzing test was performed on partially emulated firmware.
The authors mapped each firmware partition into virtual memory space separately, marked input source functions and sensitive sink functions, and generated the control-flow graph of the firmware.
They utilized a dynamic symbolic execution technique to emulate every code snippet and perform a fuzzing test to examine the memory corruption.

\subsection{SAST}
Tools for scanning source code and reporting possible security weaknesses, i.e., common security-related programming errors, include Flawfinder~\cite{flawfinder} and the Rough Auditing Tool for Security (RATS)~\cite{rats}.
Keropyan et al.~\cite{keropyan2017multiplatform} developed a platform based on a program static analysis approach for detecting use-after-free and double-free weaknesses.
To mitigate file access control vulnerabilities' harm on the Windows system, Lu et al.~\cite{lu2022static} classified file access control bugs into two types and proposed a detection method.
Alrabaee et al.~\cite{alrabaee2020binary} presented a method for performing binary analysis to address some aspects of binary code fingerprinting, such as compiler provenance, reused binary code discovery, fingerprinting of OSS packages, and authorship attribution.
Their method can be used for malware analysis, software maintenance,  software plagiarism detection, and open-source project license violation detection.

Various \ac{ML} techniques have been applied to the field of firmware vulnerability detection and prevention.
Miettinen et al.~\cite{miettinen2017iot} proposed IoT SENTINEL, a system aimed at identifying and preventing vulnerable IoT devices from affecting other devices in the network.
In their system, the random forest classification model was used to identify the device type and its firmware version.
Based on this information and by using a vulnerability database, their system restricted the communication of IoT devices if it determined that the device was vulnerable.
Lee et al.~\cite{lee2019machine} experimented with four types of \ac{ML} algorithms to identify IoT firmware information, such as manufacturer, device type, and architecture type.
Their results showed varying degrees of accuracy regarding device vendor identification.
A two-stage approach based on code similarity was proposed by Wang et al.~\cite{wang2019staged} to address the inaccuracy of vulnerability assessment using a control flow graph and simple feature matching.
Their approach utilized neural networks to analyze function similarities based on function embedding; then, local call flow graphs of the functions were produced. 

In this paper, we used some components of Karonte, a static analysis approach for analyzing embedded-device firmware by modeling and tracking multi-binary interactions, which was presented by Redini et al.~\cite{redini2020karonte}. 
Their approach propagates taint information between binaries to detect insecure interactions and identify vulnerabilities.
Karonte was improved by Cheng et al.~\cite{cheng2021automatic} who used a heuristic approach to infer taint sources rather than solving the indirect calls. 
The authors suggested automatic inference of taint sources to discover vulnerabilities in SOHO router firmware.
Liu et al.~\cite{liu2022finding} suggested a method of finding vulnerabilities in firmware, improving Karonte's means of finding connections between binaries by implementing a shared memory finder and implicit transfer method.

Hernandez et al.~\cite{hernandez2020bigmac} presented BigMAC, a framework that combined and instantiated all layers of the Android policy in a fine-grained graph.
Their model filtered out paths and types not in use on actual systems.
Using static firmware and Android domain knowledge, they extracted and recreated the security state of a running system.
The authors developed attack queries to discover sets of objects that can be influenced by untrusted applications and external peripherals.

\subsection{DAST}
Dynamic analysis is one of the main foundations of security analysis, and it relies on the ability to execute software in a controlled environment, e.g., an instrumented emulator.
Zaddach et al.~\cite{zaddach2014framework} presented Avatar, a hybrid framework that enables dynamic analysis of embedded devices, using an orchestration engine that executes an emulator along with the real hardware.
The authors demonstrated Avatar's ability to execute the firmware instructions inside the emulator (by injecting a software proxy in the embedded device) while channeling the I/O operations to the physical hardware.
Chen et al.~\cite{chen2016automated} proposed FIRMADYNE, an automated dynamic analysis system that targeted Linux-based firmware on network-connected devices.
Their system, which relies on software-based full-system emulation with an instrumented kernel, achieved the scalability necessary to analyze thousands of firmware binaries.

\subsection{Formal Validation}
Due to the complexity of the modern \ac{SoC}, security validation and verification have become challenging.
Ray et al.~\cite {ray2019formal} presented a methodology for formally verifying security flows in a commercial \ac{SoC} that involve extensive interaction between firmware and hardware.
The authors highlighted the unique challenges associated with formal security verification of these interactions and implemented their approach of property-specific abstraction and software model checking to overcome those challenges.
To address formal security verification of firmware interacting with hardware, Huang et al.~\cite{huang2018formal} presented a co-verification methodology.
They modeled hardware using the instruction-level abstraction, capturing firmware-visible behavior at the architecture level, and integrated hardware behavior with firmware into a single thread.
The co-verification with multiple firmware was formulated as a multi-threaded program verification problem for which they leveraged software verification techniques.

\subsection{Risk Assessment}
One of the most important reasons for firmware analysis is security risk assessment.
Sun et al.~\cite{sun2019tell} presented a domain-specific reverse engineering framework for embedded binary code in the \ac{IoT} control application domain; the reverse engineering outcomes can be used for firmware vulnerability assessment.
Their framework can improve understanding of cyber-physical security flaws.
A vulnerability analysis model for UEFI trust verification startup was proposed by Gu et al.~\cite{gu2020uefi}.
The authors identified the vulnerable attack paths and nodes and proposed a way of strengthening UEFI firmware's security.
Regano~\cite{regano2019expert} presented a formal risk assessment methodology for software based on AGs, in the form of an expert system for automating the protection of applications, mimicking a software security expert's decision-making process.
His system requires the program source code; in addition, the user needs to provide a list of the assets to protect, each of which must be associated with one or more high-level security requirements (e.g., confidentiality, integrity, and availability), to produce a binary of the application, hardened with the most appropriate protection technique.

To the best of our knowledge, we are the first to propose a generic AG-based framework to assess the cybersecurity risks of multi-binary firmware images.
While the aim of fuzzing, \ac{SAST}, \ac{DAST}, and formal validation is to identify weaknesses and vulnerabilities in firmware, AG-based risk assessment aims at examining the effects of known vulnerabilities in firmware images and presenting them as attack paths in an AG.

\section{Summary and Future Work} \label{sec:Futurework}
Device firmware, the software that provides low-level control for a device’s hardware, is a common target of cyberattacks.
Firmware may contain many binaries to provide the devices' functionality.
Every firmware binary that interacts with the external world is considered a potential risk.
Malicious code injected in Internet-facing binaries may propagate between processes within the firmware.
Since firmware runs different software, each of which may have different vulnerabilities and interactions with users and the external world, there may be many different possible attack paths.
AGs organize identified vulnerabilities into attack paths composed of sequences of actions an attacker can take to reach and compromise firmware.

In this paper, we used AGs for binary risk assessment, proposing MIRAGE, a static analysis framework used to identify vulnerable interactions between binaries and potential attack vectors.
MIRAGE enables the identification of the potential attack surface, interactions between the binaries, and their vulnerabilities.
To evaluate the proposed framework, we collected a dataset of 703 firmware images and generated AGs for them. 
We presented the average number of vulnerable binaries in each firmware, proposed a model for determining the impact and likelihood of each binary, and listed the binaries in our dataset with the greatest risk.

Firmware risk assessment and the identification of risky binaries are essential for improving the security of today’s widely used \ac{IoT} devices and the performance of the security teams responsible for defending them.
Security practitioners can use the proposed MIRAGE framework and its risk assessment model to prioritize the firmware risks and binary patching.

Future research could focus on improving the process of finding each binary's version and identifying additional connections between the binaries by examining other ways binaries communicate.
Since complex software systems also have Internet-facing components and complex interactions between components that have various vulnerabilities, in future work, the proposed framework could be applied to different software systems.
Research could also examine the AG visualization required by various stakeholders in the security domain and the generation of alerts based on risks identified by the AG.
Another future research direction is to explore the use of AGs to assess security risk in other areas, including the dynamic environments of containers and connected vehicles.

\bibliographystyle{IEEEtran} 
\bibliography{References}

\appendix
\section{Acronyms}
\begin{acronym}
    \acro{AG}{attack graph}
    \acro{CI/CD}{continuous integration and continuous delivery}
    \acro{CISA}{Cybersecurity \& Infrasructure Security Agency}
    \acro{CVE}{common vulnerabilities and exposures}
    \acro{DAST}{dynamic application security testing}
    \acro{EPSS}{exploit prediction scoring system}
    \acro{ICS}{industrial control system}
    \acro{IDS}{intrusion detection system}
    \acro{IoT}{Internet of things}
    \acro{IRs}{interaction rules}
    \acro{LAG}{logical attack graph}
    \acro{ML}{machine learning}
    \acro{NVD}{National Vulnerability Database}
    \acro{OS}{operating system}
    \acro{OSS}{open-source software}
    \acro{SAST}{static application security testing}
    \acro{SDKs}{software development kits}
    \acro{SoC}{system-on-chip}
    \acro{TTPs}{tactics, techniques, and procedures}
\end{acronym}

\end{document}